\documentclass{aa} 

\usepackage{graphicx}
\usepackage{txfonts}
\usepackage{lscape}
\usepackage{color}
\usepackage{soul}
\usepackage[english]{babel}
\begin{document} 

   \titlerunning{He enhancement among M3's HB stars}


   \title{Level of helium enhancement\\ among M3's horizontal branch stars}

   \author{A.~A.~R. Valcarce\inst{1,2,3}
          \and
          M. Catelan\inst{1,2,3}
          \and
	  J. Alonso-Garc\'ia\inst{4,3}
	  \and
	  R. Contreras Ramos\inst{1,3}
	  \and
	  S. Alves\inst{1,3}
          }

   \institute{
   	    Pontificia Universidad Cat\'olica de Chile, Instituto de Astrof\'isica, Facultad de F\'isica, Av. Vicu\~na Mackenna 4860, 782-0436 Macul, Santiago, Chile
	 \and
            Pontificia Universidad Cat\'olica de Chile, Centro de Astroingenier\'ia, Av. Vicu\~na Mackenna 4860, 782-0436 Macul, Santiago, Chile
         \and
	    Millennium Institute of Astrophysics, Santiago, Chile
	 \and
	    Universidad de Antofagasta, Unidad de Astronomía, Facultad Cs. Básicas,  Av. U. de Antofagasta 02800, Antofagasta, Chile
             }

   \date{Received April 22, 2015; accepted January 25, 2016}

 
  \abstract
   {The color and luminosity distribution of horizontal branch (HB) stars in globular clusters (GCs) are sensitive probes of the original helium abundances of those clusters. In this sense, recently the distributions of HB stars in GC color-magnitude diagrams (CMDs) have been extensively used as indicators of possible variations in the helium content $Y$ among the different generations of stars within individual GCs. However, recent analyses based on visual and near-ultraviolet (UV) CMDs have provided conflicting results.}
   {To clarify the situation, we address the optimum ranges of applicability (in terms of the $T_{\rm eff}$ range covered by the HB stars) for visual and near-UV CMDs, as far as application of this ``HB $Y$ test'' goes.}
   {We considered both Str\"omgren and {\em Hubble Space Telescope} (HST) bandpasses. In particular, we focus on the F336W filter of the HST, but also discuss several bluer UV bandpasses, such as F160BW, F255W, and F300W. Using the Princeton-Goddard-PUC (PGPUC) code, we computed a large set of zero-age HB (ZAHB) loci and HB evolutionary models for masses ranging from $M_{\rm HB}=0.582$ to $0.800 \, M_\odot$, assuming an initial helium abundance $Y=0.246$, 0.256, and 0.266, with a global metallicity $Z=0.001$. The results of these calculations were compared against the observations of M3 (NGC\,5272), with special attention on the $y$ vs. $(b-y)$ and F336W vs. (F336W$-$F555W) CMDs.}
   {Our results indicate that, from an evolutionary perspective, the distributions of HB stars in the $y$ vs. $(b-y)$ plane can be a reliable indicator of the He content in cool blue HB (BHB) stars, particularly when a differential comparison between blue and red HB stars is carried out in the range $T_{\rm eff}\lesssim 8\,300$~K. Conversely, we demonstrate that CMDs using the F336W filter have a much less straightforward interpretation at the cool end of the BHB because the distributions of HB stars in the F336W vs. (F336W$-$F555W) plane, for instance, are affected by a triple degeneracy effect. In other words, the position of an HB star in such a CMD is exactly the same for a given chemical composition for multiple combinations of the parameters $Y$, $M_{\rm HB}$, and age along the HB evolutionary track. Other HST UV filters do not appear to be as severely affected by this degeneracy effect, to which visual bandpasses are also immune. On the other hand, such near-UV CMDs can be extremely useful for the hottest stars along the cool BHB end.
   }
   {Based on a reanalysis of the distribution of HB stars in the $y$ vs. $(b-y)$ plane, we find that the coolest BHB stars in M3 (i.e., those with $T_{\rm eff}<8\,300\,K$) are very likely enhanced in helium by $\Delta Y\approx 0.01$, compared with the red HB stars in the same cluster. Using near-UV HST photometry, on the other hand, we find evidence of a progressive increase in $Y$ with increasing temperature, reaching $\Delta Y \approx 0.02$ at $T_{\rm eff} \approx 10\,900$~K.
   }


   \maketitle
%

\section{Introduction}

Globular star clusters (GCs) are far from being the simple stellar populations that they were once thought to be. The presence of multiple populations, corresponding to different episodes of star formation within individual GCs, is now routinely revealed by photometry and spectroscopy alike. One important open problem is the degree to which the different populations may differ in their helium content, and how any internal spread in the helium abundance $Y$ may differ from one cluster to the next. 

The discovery of multiple main sequences (MSs) in NGC~2808 \citep{Norris2004, DAntona_etal2005, Piotto_etal2007} and $\omega$~Centauri \citep[NGC~5139;][]{Bedin_etal2004, Bellini_etal2009, Bellini_etal2010} has reopened the question about the initial helium abundance of the GC stars \citep{van_den_Bergh1965, van_den_Bergh1967, Sandage_Wildey1967, Hartwick1968}. Remarkably, NGC~2808 shows three discrete MS components, strongly suggesting large He variations from one such component to the next \citep{Piotto_etal2007}, even though the cluster does not have a large spread in metallicity \citep{Gratton_etal2011}. In like vein, a bluer MS component in $\omega$~Cen also appears to require a much higher He abundance than would have been implied by their difference in metallicity \citep{Piotto_etal2005}. Even though it is believed that most of the stellar populations of GCs follow a helium-to-metal enrichment law \citep[$Y\approx Y_p + \Delta Y/\Delta Z \times Z$, with $\Delta Y/\Delta Z\approx 2.0$;][]{Renzini1994, Salaris_etal2004, Valcarce_etal2013}, the discrete main sequences and the different chemical compositions observed among stars in the same GC suggested the existence of helium-rich stars in every Milky Way GC \citep{Gratton_etal2012}. Nevertheless, the fraction of these He-rich stars in GCs and the amount of helium enrichment in each cluster are still a matter of debate.

Because the surface helium abundance can only be measured spectroscopically in stars with an effective temperature $T_{\rm eff}\gtrsim 8\,000$~K, only GCs presenting relatively blue horizontal branch (HB) extensions can harbor stars with sufficiently high temperatures to enable direct spectroscopic estimations of the helium abundances. However, these stars have passed through the whole MS, red giant branch (RGB), and pre-HB phases, which has modified their surface chemical composition through the effects of internal mixing \citep[e.g.,][]{Sweigart_Mengel1979,VandenBerg_Smith1988}, and they might also have been affected by stellar rotation \citep[e.g.,][]{Mengel_Gross1976}, mass loss \citep[e.g.,][]{Peterson1982, DeLaReza_etal1996}, and/or interaction with other bodies \citep[planets or stars, ][]{Soker1998, Siess_Livio1999}, among other effects \citep[see][]{Catelan2009}.

Even for these hot HB stars, spectroscopic measurements of helium unfortunately remain complex because these lines are weak and a high signal-to-noise ratio (S/N) is required to derive accurate results. Moreover, it is well established that GC HB stars with $T_{\rm eff}\gtrsim 11\,500$ K show higher metal and lower helium abundances than all the other stars of the same GC. This is caused by the effects of metal levitation and He sedimentation \citep{Grundahl_etal1999,Behr2003}, which limit the range of temperatures for measuring helium to a window of $\Delta T_{\rm eff} \approx 3500$ K. Only recently, \citet{Villanova_etal2012} and \citet{Marino_etal2013} have obtained high-precision spectral measurements of helium abundances in the stars of M4 (NGC~6121) and NGC~2808, respectively. We still lack a complete understanding, however, of how these photospheric abundances are related to the initial abundances after the whole previous evolution, given in particular their dependence on details of the first dredge-up episode at the base of the RGB. This phenomenon is only poorly described by canonical evolutionary models without extra mixing \citep[e.g.,][]{Gratton_etal2000}. Other spectroscopic helium abundance studies include those of \citet{Dupree_etal2011} and \citet{Pasquini_etal2011} for RGB stars in $\omega$~Cen and NGC~2808, respectively. These authors used chromospheric helium lines (which have more traditionally been used as diagnostics of mass loss in RGB stars) that require a careful treatment of the stellar atmosphere \citep{Dupree_Avrett2013}. These difficulties have favored indirect methods to be proposed as probes of the He abundance in GC stars.

For this purpose, the effects of different initial helium abundances on color-magnitude diagrams (CMDs) have been known since the late 1960s (\citealt{Simoda_Iben1968,Aizenman_etal1969, Iben1974, Simoda_Iben1970}; see also \citealt{Valcarce_etal2012}, for a recent summary). These studies revealed that, even though these effects are present in all the evolutionary sequences along the CMD, the effect of He is amplified in the HB phase and is also easier to detect because of the high luminosities of HB stars \citep{Faulkner_Iben1966, Iben_Faulkner1968, Iben_Rood1970, Sweigart1987}. Specifically, an increase in $Y$ increases the luminosity of all HB stars with $T_{\rm eff}\lesssim 16\,000$ K, but it also decreases the luminosity of these stars for $T_{\rm eff}\gtrsim 16\,000$ K. These effects are due to different efficiencies of the H-burning shell and to the differences in the helium-core mass \citep[$M_{\rm cHe}$;][]{Valcarce_etal2012} with changing $Y$.

However, depending on the set of filters used to transform from the theoretical to the observational plane, these effects can become more or less evident in the HB part of the CMD, and thus more or less easily detectable. In this paper, we explore the effects on the HB of using two different sets of filters, Str\"omgren and {\em Hubble Space Telescope} (WFPC2), covering the visual and near-UV regimes, for setting constraints on the initial helium abundances of the different stellar populations in GCs.

\section{Estimating $\Delta Y$ from HB populations}
\label{Methodology}

Since the discovery of multiple populations in GCs, two main approaches have been used to estimate differences in the initial helium abundances in the HB population. The first approach involves the computation of synthetic HBs, which can then be compared to the distribution of stars observed along the HB \citep[e.g.,][]{Rood1973, Catelan_etal1998, DAntona_etal2002, Caloi_DAntona2008, DAntona_Caloi2008, Dalessandro_etal2011, Dalessandro_etal2013, Salaris_etal2013, Lei_etal2013, Lei_etal2013a}. In the second approach, reference evolutionary loci are directly overplotted on the empirical CMD data \citep[\citealt{Catelan_etal2009}, hereafter C09;][hereafter V14]{Brown_etal2010, Valcarce_etal2014}. 

Both approaches are based on extensive calculations of evolutionary models for GC stars, calculated according to the chemical composition of the GC under study. More specifically, the complete evolution of a star from the main sequence to the ZAHB is calculated, most often without mass loss, with the resulting maximum ZAHB mass depending on the GC properties (e.g., for 13~Gyr, $Z=0.001$, and $Y=0.245$, the maximum ZAHB mass is $0.810\,M_\odot$). This reference ZAHB model is then used as a starting point to create i)~the ZAHB locus for all stars with the same $M_{\rm cHe}$ and initial chemical composition (i.e., all stars born at the same time), and ii)~the initial model for HB stars at a given total mass ($M_{\rm HB}$) with the same $M_{\rm cHe}$ and different envelope mass \citep[i.e., due to mass loss during the RGB evolution;][]{Serenelli_Weiss2005}. Then we compute for each $M_{\rm HB}$ value the evolution from the ZAHB to the helium core exhaustion (also called terminal-age HB, or TAHB). The models are then transformed to the empirical CMD planes, using suitable bolometric corrections and color transformations. 

In the case of synthetic HB populations, the color and absolute magnitude distributions of stars along the HB are then simulated, using Monte Carlo methods. This requires an assumption on the mass-loss distribution on the RGB, as this defines the resulting distribution in masses on the HB phase proper. Most commonly, mass-loss efficiency is treated as a free parameter that is constrained a posteriori to match the observations. Stars are thus fed into the evolutionary tracks of different masses according to the expected evolutionary speed along the HB phase. Photometric errors are then added to the photometric data for the ``synthetic stars'' to more realistically compare them with observational datasets. Dozens of simulations are typically run, for which some or all of the free parameters are varied to find the distribution that most closely describes the empirical data. Some of these comparisons are made using color and magnitude histograms, but in other cases, only a visual inspection is used, which can sometimes lead to misinterpretations. 

While the reference evolutionary loci used in the second approach can also be obtained from synthetic HB models, they are more directly retrieved from the HB evolutionary tracks proper. We have already encountered two such reference loci, namely the ZAHB and TAHB. Additional loci may easily be defined. They represent the fraction of the entire HB lifetime that a given HB has already gone through. For instance, C09 defined the middle-age HB (MAHB) locus that closely corresponds to stars that are half-way through their HB evolution, and also the 90AHB, which corresponds to stars that have completed 90\% of their lives as HB stars.\footnote{C09 point out that the MAHB (originally defined as the HB ridgeline in the CMD) and 50AHB (defined in complete analogy with the 90AHB) loci may differ slightly, but here we define the MAHB as being synonymous with 50AHB.}

If a given GC contains a single population of stars, one would then expect to find $\approx50\%$ of all HB stars between the ZAHB and MAHB, $\approx40\%$ between the MAHB and 90AHB, $\approx10\%$ between the 90AHB and TAHB, etc. (see also C09). However, if there are several populations of HB stars with similar temperatures but different initial helium abundances, these percentages should change as a result of the increase in luminosity of HB stars with the increase of the initial helium abundance. In particular, if there is a spread in the initial helium abundance $\Delta Y=0.01$ in a GC and models for a single $Y$ are compared to photometric data, the percentages of stars located between the MAHB and 90AHB loci should be larger than expected on the basis of single-$Y$ models. Thus the evolutionary loci method, like the synthetic HB populations method, can be a powerful probe of the presence of populations with different He abundances that co-exist in a GC.

While HB stars with different $Y$ values can coexist over a given temperature range, theoretical arguments \citep{Sweigart_Gross1976} as well as recent spectroscopic observations \citep{Marino_etal2011, Villanova_etal2012, Marino_etal2013} alike suggest a tendency for blue HB (BHB) stars to be preferentially He-enhanced, compared with red HB (RHB) stars in the same cluster.\footnote{BHB and RHB stars are terms that are often used to refer to stars bluer and redder, respectively, than the RR Lyrae instability strip, respectively.} As a consequence, if a spread in He is indeed present within a GC with both blue and red HB components, one would expect the BHB to be noticeably brighter than the RHB in the same cluster. The size of the effect, over the otherwise rather flat part of the HB, is about $0.03$~mag in Johnson $V$ or Str\"omgren $y$ for a difference in $Y$ of 0.01. We call this comparison between the brightness of RHB and BHB stars using the ZAHB and other reference loci the HB $Y$ test.

\begin{figure*}[t]
\includegraphics[height=12.0cm, trim=0.0cm 5.8cm 0cm 1cm]{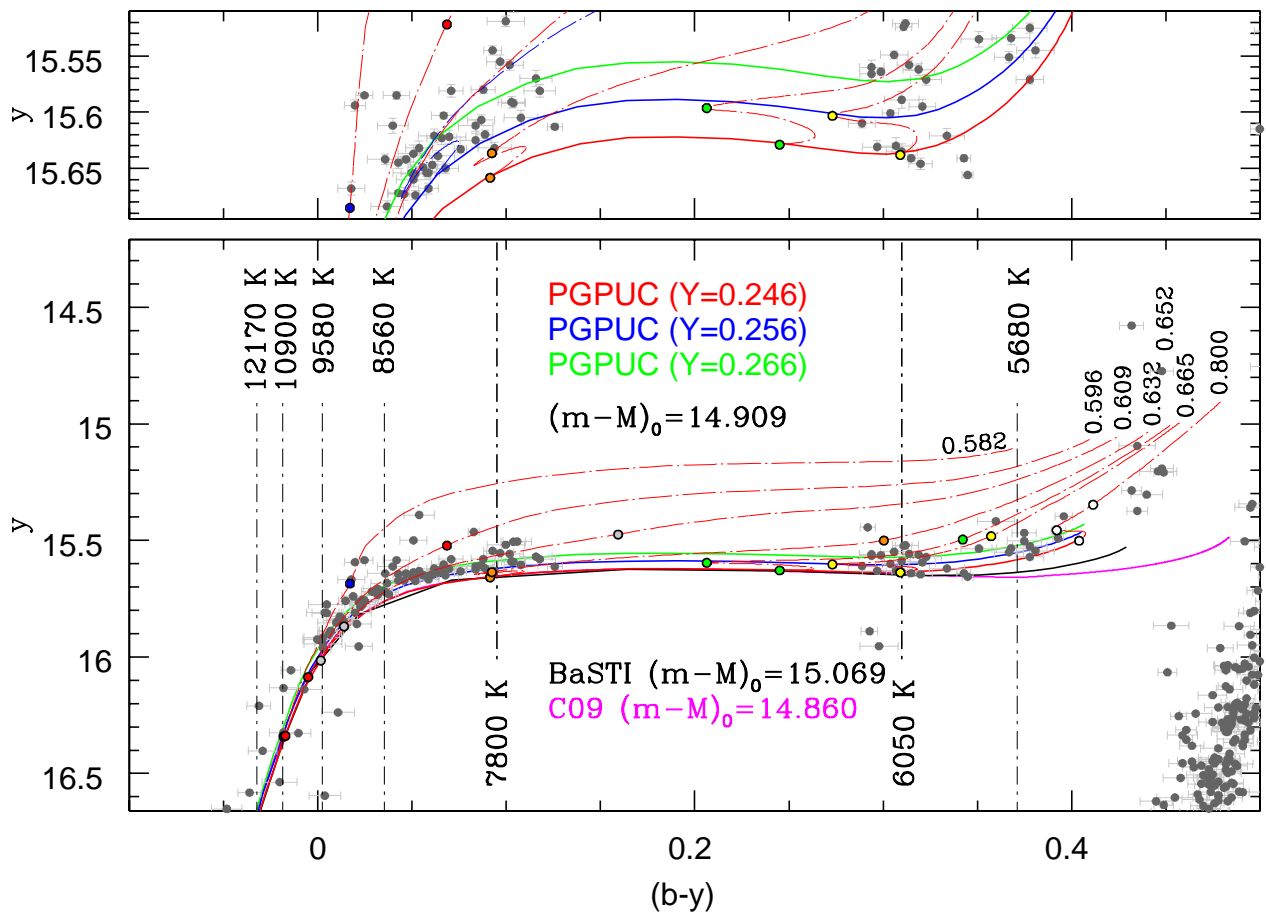}
\caption{M3 HB stars compared to theoretical HB models with $Z=0.001$ in the Str\"omgren $y$ vs. $(b-y)$ plane. The upper panel is a zoom-in around the ZAHB locus. PGPUC ZAHB loci are shown for helium abundances $Y=0.246$ (red line), $0.256$ (blue line, shown in the upper panel only, for clarity), and $0.266$ (green line). C09 and BaSTI ZAHB loci for $Y=0.246$ are also shown as magenta and black lines, respectively, with different distance moduli in order to match the PGPUC ZAHB locus with $Y=0.246$. Theoretical HB evolutionary tracks for $Y=0.246$ are shown as red dot-dashed lines for $M_{\rm HB}=0.582$, 0.596, 0.609, 0.632, 0.652, 0.665, and $0.800\, M_\odot$, which have been marked with circles at the ZAHB, MAHB, and 90AHB (colors depend on the HB mass, for clarity). An additional HB track for $Y=0.256$ and $M_{\rm HB}=0.582\,M_\odot$ is also shown in the top panel (blue dot-dashed lines with pink circles). Vertical dot-dashed lines indicate some $T_{\rm eff}$ values at the ZAHB locus for $Y=0.246$. The distance modulus was selected as described in Sect.~\ref{SectDistMod} and Fig.~\ref{FigHBDmod}. The data come from C09.}
\label{FigHBTracksStr}
\end{figure*}

It is worth mentioning that stars with high helium abundances could in principle also have suffered an increase in $Z$ or C+N+O, as suggested recently by \citet{Jang_etal2014} in their proposed explanation of the Oosterhoff dichotomy observed in Milky Way GCs (\citealt{Oosterhoff1939}; see also \citealt{Catelan2009}, for a recent review and references). This can also favor helium-normal and helium-rich stars sharing similar colors on the HB of a given GC, and therefore serves as a warning against solely using color information as an indicator of He spreads among GC stars.\footnote{There are other possibilities for He-rich and He-poor stars to share similar colors on the HB. For example, stars with the exact same $Z$ and C+N+O but different mass-loss histories may very well have exactly the same ZAHB temperature, even though their luminosities may be vastly different.} In particular, the percentages between ZAHB, MAHB, 90AHB, and TAHB loci are expected to change, when compared to the expectations for a single stellar population because of the difference in luminosity that is brought about by the presence of a helium spread.

\section{The M3 case revisited}
\label{ZAHBfilters}

Here we consider M3 (NGC~5272) because this cluster has been the subject of numerous recent studies of the HB population \citep[C09;][hereafter D13]{Dalessandro_etal2013}, without a clear consensus as yet on whether there is a He spread, and if so, by how much. One of our main goals is to explain the reasons for seemingly conflicting results based on similar samples of HB stars in this one cluster. 

In the following, we assume that there is no difference in either $Z$ or C+N+O among stars in M3 \citep[$\sigma_{\rm [Fe/H]}=0.03$;][]{Sneden_etal2004} so that we can compare the expected and observed luminosity differences using different sets of filters.

\begin{figure*}[t]
\includegraphics[height=12.0cm, trim=0.0cm 5.8cm 0cm 1cm]{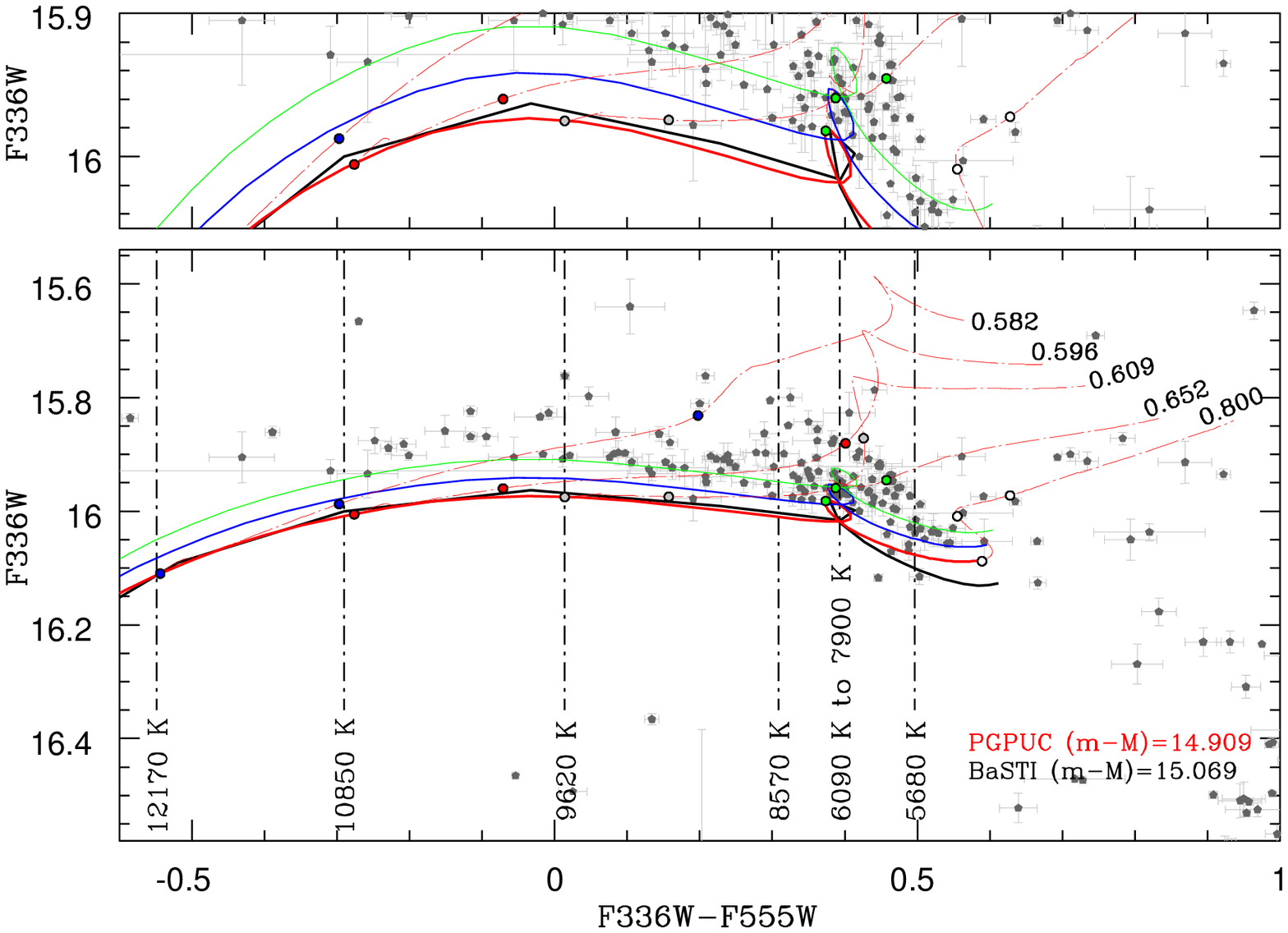}
\caption{Same as Fig.~\ref{FigHBTracksStr}, but for the HST F336W vs. (F336W$-$F555W) plane. Theoretical models for $M_{\rm HB}=0.665$ and $0.632$ $M_\odot$ are not shown here for clarity (see Fig.~\ref{FigHBTracksUVzoom}). The observational data are the same as in D13, as kindly provided by E. Dalessandro (2014, priv. comm.).}
\label{FigHBTracksUV}
\end{figure*}

Figure~\ref{FigHBTracksStr} shows the HB population of M3 on a CMD based on the $y$ vs. $y-b$ filters of the Str\"omgren system. Similarly, Fig.~\ref{FigHBTracksUV} shows the $F336W$ vs. $F336W-F555W$ CMD, using the HST Wide-Field and Planetary Camera 2 (WFPC2) visible (F555W) and near-UV (F336W) filters, respectively. M3 observational data in the Str\"omgren filters are taken from \citet{Grundahl_etal1998, Grundahl_etal1999} as described in C09 with a total of 138 non-variable HB stars, while data in the HST filters were kindly provided by E. Dalessandro (2014, priv. comm.) with a total of 155 non-variable HB stars. In these figures are also plotted theoretical ZAHB loci for a chemical composition according to the M3 metallicity ($Z=0.001$ and $[\alpha/{\rm Fe}]=0.30$; see C09) and three initial helium abundances $Y=0.246$, 0.256, and 0.266 as red, blue, and green lines, respectively. HB evolutionary tracks for $Y=0.246$ are also plotted as red dot-dashed lines for some masses ($M_{\rm HB}=0.582$, 0.596, 0.609, 0.632, 0.652, 0.665, and $0.800 \, M_\odot$). These are marked with circles at three stages of their evolution~-- ZAHB, MAHB, and 90AHB~-- with different colors being used for different $M_{\rm HB}$ values. These figures are explained in more detail below.

\subsection{Previous studies}
The morphology of the HB population of M3 (${\rm [Fe/H]}=-1.50$) was studied in D13 together with that of two other GCs, namely M13~= NGC\,6205 and M79~= NGC\,1904, which have very similar [Fe/H] ratios \citep[$-1.53$ and $-1.60$, respectively;][2010 edition]{Harris1996}. D13 compared the results of synthetic HB calculations to HST near-UV observations that were obtained with WFPC2. These simulations require four free parameters (the minimum $Y$ value, the range of helium abundances, the mean value of the mass lost along the RGB, and the spread around this mean value) for an assumed age of 12~Gyr. The amount of mass lost on the RGB is also expected to be a function of $Y$ \citep[e.g.,][]{Catelan_deFreitasPacheco1993}. Clearly, the number of possible parameter combinations and the resulting synthetic distributions is very large, even though in D13 it is only possible to examine the best-fitting simulation for each GC. For M3, these synthetic HB calculations suggest, according to D13, that HB stars with $0.2\lesssim({\rm F336W}-{\rm F555W})\lesssim0.5$ have a dispersion in $Y$ of $0.02$, but the dispersion in $Y$ advocated by D13 for M13 is only 0.01 over the same color range. Although small, the presence of such a difference is noteworthy, given the similar brightness extensions in F336W of the HB populations in both clusters. This difference might plausibly be traced back to both clusters' strikingly different HB morphologies: M3 and M13 constitute a ``classical'' second-parameter case, and one of the first for which chemical anomalies, as opposed to age, were suggested as the explanation \citep{Catelan_deFreitasPacheco1995}.

The population of HB stars was studied by C09 using the visual filters of the Str\"omgren system, and the resulting CMDs are compared to different reference evolutionary loci (ZAHB, MAHB, 90AHB, and TAHB; see Sect.~\ref{Methodology}). As can be seen from Fig.~2 in C09, the evolution of HB stars in these filters ($M_y$, $M_b$, and $M_v$) implies that, in the canonical scenario, brighter stars (at a given temperature) are also more evolved stars, at least for $T_{\rm eff}\lesssim9\,500$~K. However, since an increase in $Y$ also induces an increase in the HB luminosity at similar temperatures \citep[see, e.g.,][]{Sweigart1987}, an increase in $Y$ can be mimicked by different ages along the HB evolutionary tracks ($t_{\rm HB}$), and vice-versa. However, as mentioned before, theoretical evolutionary models predict that 50\% of all HB stars must be found between the ZAHB and MAHB loci, or, similarly, between the ZAHB locus with primordial $Y$ value and the ZAHB locus with a $Y$ value increased by 0.01, as the latter's ZAHB closely overlaps the former's MAHB. Based on this, the main result in C09 is that any helium spread in M3 must accordingly be $\Delta Y \lesssim 0.01$.

In summary, the proposed level of He enhancement among M3 HB stars proposed by D13 is in conflict with the smaller one suggested in C09 for stars over the same range in effective temperature (i.e., basically for the same stars). Below we address some possible reasons for these conflicting results.

\subsection{Comparing models in the visual and near-UV}

\begin{figure}
\includegraphics[height=9.15cm, trim=0.0cm 0cm 0cm 0.0cm]{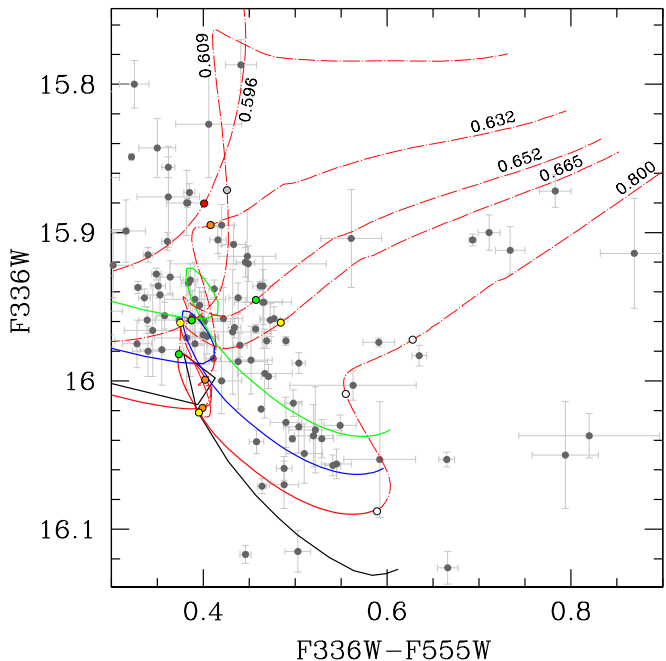}
\caption{Zoom-in of Fig.~\ref{FigHBTracksUV} around the {\em loop} shown by the ZAHB loci. Here are shown all the computed HB evolutionary tracks. Note the difficulty disentangling a variation in $Y$ from mass and evolutionary state, at colors around $({\rm F336W}-{\rm F555W}) \approx 0.4$. }
\label{FigHBTracksUVzoom}
\end{figure}

\begin{figure*}
\includegraphics[height=18cm, trim=0cm 0cm 0cm 0cm]{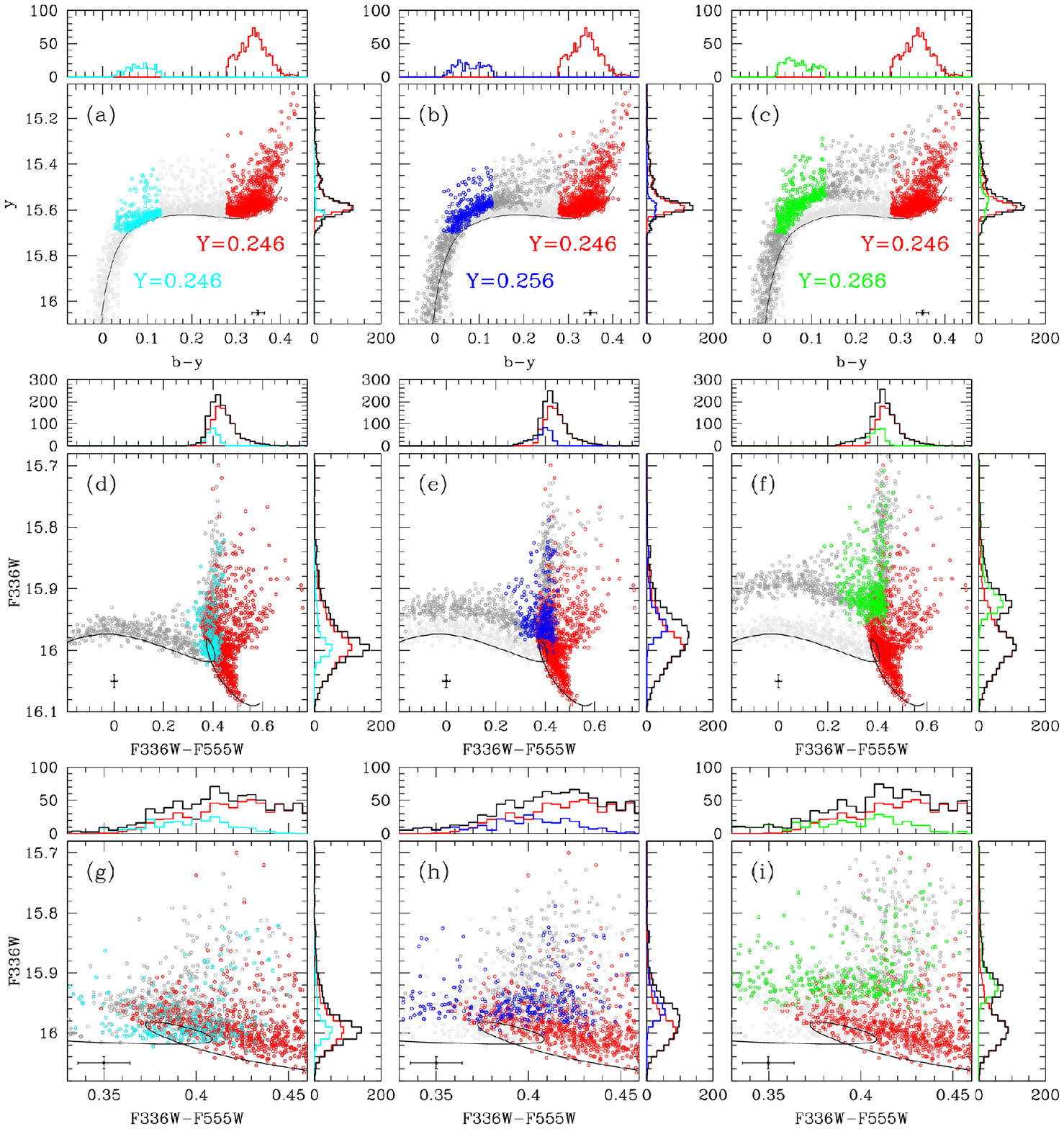}
\caption{Synthetic CMDs at the HB level from models with $Z=0.001$ in the filters of interest of the Str\"omgren (top panels) and WFPC2 (middle and bottom panels) systems. The ZAHB locus for $Y=0.246$ is represented as a black line. A uniform mass deviate ranging from $0.595$ to $0.705 \, M_\odot$ with 2100 stars is used to represent a helium-normal population ($Y=0.246$, light gray circles). Helium-rich populations ($Y=0.256, 0.266$) are created with 1000 stars with masses ranging from $0.595$ to $0.645 \, M_\odot$ (dark gray circles). Stars identified as RHB stars with $Y=0.246$ are plotted as red circles. BHB stars are plotted in cyan for $Y=0.246$, blue for $Y=0.256$, and green for $Y=0.266$. In these simulations, photometric errors are included, following the photometric data, at the level of 0.01~mag in each bandpass. Panels g, h, and i are zoom-ins of panels d, e, and f, respectively. Each panel is accompanied by histograms showing the distribution in color (top) and magnitude (right) of the different components.}
\label{FigSyntheticHBs}
\end{figure*}

To look further into the reasons for the noted discrepancy between the results of C09 and D13, we show in Figs.~\ref{FigHBTracksStr} and \ref{FigHBTracksUV} HB evolutionary tracks and reference loci in CMDs based on the same filters as used in C09 and D13 together with the observational datasets used by these authors. Figure~\ref{FigHBTracksUVzoom} is a zoom-in of Fig.~\ref{FigHBTracksUV} in the color range $0.2\lesssim({\rm F336W}-{\rm F555W})\lesssim0.8$. In these figures we show HB evolutionary tracks (red dot-dashed lines) with masses ranging from $M_{\rm HB}=0.800$ down to $0.582 \, M_\odot$, with a progenitor mass of $0.800 \, M_\odot$, $Y=0.246$, $Z=0.001$, and $[\alpha/{\rm Fe}]=+0.3$. To show the different evolutionary stages, each HB track has been marked with colored circles at the positions in their evolution corresponding to the ZAHB, MAHB, and 90AHB, the colors of the circles depending on the stellar mass. The distance modulus $(m-M)=14.909$ is selected to reproduce the expected proportions of stars according to evolutionary stage along the HB phase (see Sect.~\ref{SectDistMod}).\footnote{Even though the distance modulus is lower than reported in the literature \citep[e.g., ][]{VandenBerg_etal2013}, this does not affect the main results of our paper.} These figures also show the BaSTI \citep{Pietrinferni_etal2004} and C09 ZAHB loci (in black and magenta colors, respectively), shifted according to the indicated distance moduli in order to match the PGPUC ZAHB locus for $Y=0.246$ (red line). Note that the different required shifts, hence distance moduli, are due to differences in the adopted chemical compositions (in C09, the $Y=0.230$ ZAHB locus is used), bolometric corrections (C09 and PGPUC models use \citeauthor{Clem_etal2004} 2004, whereas BaSTI models use \citeauthor{Castelli_Kurucz2004} 2004), and/or input physics \citep[more details can be found in Sect.~2.7 of ][]{Valcarce_etal2012}. Blue and green lines represent the PGPUC ZAHB loci for $Y=0.256$ and $Y=0.266$, respectively. All these ZAHB loci are based on the same chemical composition mentioned before, except for the C09 model, whose higher metallicity ($Z=0.002$) explains its further extension toward red colors, compared to the other ZAHB loci shown. Finally, vertical dash-dotted lines schematically indicate some specific $T_{\rm eff}$ values (based on the ZAHB loci), as were used to compare different sets of filters. For the sake of clarity, evolutionary tracks for 0.665 and $0.632 \, M_\odot$ are not shown in Fig.~\ref{FigHBTracksUV}, and neither is the C09 ZAHB locus.

A quick inspection of these figures reveals that the $Y$-$t_{\rm HB}$ degeneracy existing in the HB with the filters used in Fig.~\ref{FigHBTracksStr} ($y$ vs. $b-y$ plane) can induce a misinterpretation not larger than $\Delta Y=0.01$ (compare the blue line in the upper panel with the overplotted evolutionary tracks) because roughly 50\% of the HB stellar population with $Y=0.246$ must be brighter than the ZAHB locus with $Y=0.256$ in the temperature range $5\,700\,{\rm K}\lesssim T_{\rm eff}\lesssim 8\,300\,{\rm K}$, regardless of the existence of a helium spread.

However, the filters used in Fig.~\ref{FigHBTracksUV} (F336W vs. ${\rm F336W}-{\rm F555W}$ plane) present a $Y - t_{\rm HB} - M_{\rm HB}$ triple degeneracy at the HB level. Specifically, from red to blue colors, the ZAHB locus starts with $0.800\, M_\odot$ at a color around $0.6$, and when the mass is decreased to $\sim0.665\, M_\odot$, the color is $0.4$~--, but for lower masses, the ZAHB locus begins to develop a {\em loop}. In other words, a further decrease in mass, instead of producing a bluer structure with roughly the same brightness as in Fig.~\ref{FigHBTracksStr}, gives rise to a redder and brighter CMD position. A peak in brightness is reached for a ZAHB mass of $0.652\, M_\odot$. When the mass is decreased even more, one finally returns to a color of 0.4, which occurs for $M_{\rm HB}=0.632\, M_\odot$. Finally, the usual monotonic behavior is again recovered for even lower masses or higher temperatures. 

This non-monotonic behavior and the resulting triple degeneracy in HB parameters seriously hinders the interpretation of photometric observations with the F336W filter, particularly in and around the temperature range where the loop occurs, as indicated in Fig.~\ref{FigHBTracksUV}~-- that is, between about $5\,800\,{\rm K}$ and $8\,300\,{\rm K}$. Unfortunately, many M3 HB stars live at this temperature range, including all of its RR Lyrae and the redder of its BHB stars. Note that $72\pm7\%$ of the M3 total HB population has $T_{\rm eff}<8\,300\,{\rm K}$. In addition, as a result of photometric errors (see the error bars plotted in Fig.~\ref{FigHBTracksUVzoom}), this effect potentially affects the evolutionary interpretation of not only those few stars that are located within the exact boundaries of these loops, but also of stars in their immediate neighborhood, in the color range $0.34 \lesssim({\rm F336W}-{\rm F555W})\lesssim 0.45$, which comprises a significant fraction ($28\pm5\%$) of the non-variable HB stars in M3. Note that this fraction increases to $55\pm 6\%$ when RR Lyrae stars occupying similar colors are also included \citep[][their Table~1 for $r<50''$]{Catelan_etal2001}. 

The preceding analysis can be further explored using synthetic HB models, as shown in Fig.~\ref{FigSyntheticHBs}, where synthetic CMDs at the HB level are shown in the Str\"omgren and WFPC2 systems. In each case, a uniform mass deviate was used with $200$ HB stars per $0.01 \, M_\odot$, leading to $T_{\rm eff}$ and $\log L$ values for each star which are then transformed into the observational planes using the \citet{Clem_etal2004} and \citet{Castelli_Kurucz2004} bolometric corrections for $y$ vs. $(b-y)$ and F336W vs. $({\rm F336W}-{\rm F555W})$, respectively. The photometric uncertainties, in all bandpasses, are at the level of 0.01~mag on average and were also modeled after normal deviates. The left panels (a, d, and g) show a stellar population with $Y=0.246$ with masses between $0.595$ and $0.705 \, M_\odot$ (light gray circles), and panels b and e (c and f) show a helium-enriched stellar population with $Y=0.256$ ($Y=0.266$) with masses between $0.595$ to $0.645\, M_\odot$ (dark gray circles).

To show how the high percentage of BHB and RHB stars around the aforementioned loop feature mix in magnitude and color for (F336W$-$F555W), RHB stars are cleanly selected in the Str\"omgren system as those with $0.28\le(b-y)$, and BHB as those with $0.02\le(b-y)\le0.13$ and $y\le15.7$. The same stars are then accordingly labeled in the CMDs that use the WFPC2 filters. The RHB populations with $Y=0.246$ are shown as red circles, while BHB populations with $Y=0.246$, 0.256, and 0.266 are shown as cyan, blue, and green circles, respectively. As expected, in the Str\"omgren $y$ vs. $(b-y)$ CMD, the color distribution separates almost completely BHB stars with $Y=0.256$ or 0.266 from RHB stars with $Y=0.246$, except for a small fraction (of about 5\% or less) corresponding to highly evolved BHB stars ($t_{\rm HB}>90\%$). The upper panels show that the $y$ vs. $(b-y)$ plane can be used reliably only for $(b-y)\gtrsim0.06$ ($T_{\rm eff}\lesssim8\,300$K) because of the observational errors. Thus, unless the photometric errors are smaller than indicated and/or these errors are properly taken into account in the analysis, no safe constraints can be placed on the level of He enhancement for such hot HB stars based solely (as in C09) on the $y$ vs. $(b-y)$ CMD.

For the F336W vs. $({\rm F336W}-{\rm F555W})$ plane, we find, as expected, that the color distribution around the loop cannot be used to separate BHB from RHB stars in the range $0.34 \lesssim({\rm F336W}-{\rm F555W})\lesssim 0.45$ ($8\,300\gtrsim T_{\rm eff}[{\rm K}]\gtrsim5\,800$). In turn, luminosity distributions can be used to identify the existence of a helium-enriched BHB population, but only if RHB and BHB populations have similar numbers of stars populating the relevant color range and if the statistics is high. Since it is well known that helium-enriched populations often comprise only a fraction of the helium-normal population in GCs, a reliable characterization of the stars around the loop feature in terms of their He enhancement levels is very challenging.

\begin{figure*}
\includegraphics[height=18cm, trim=0.0cm 0cm 0cm 0.0cm]{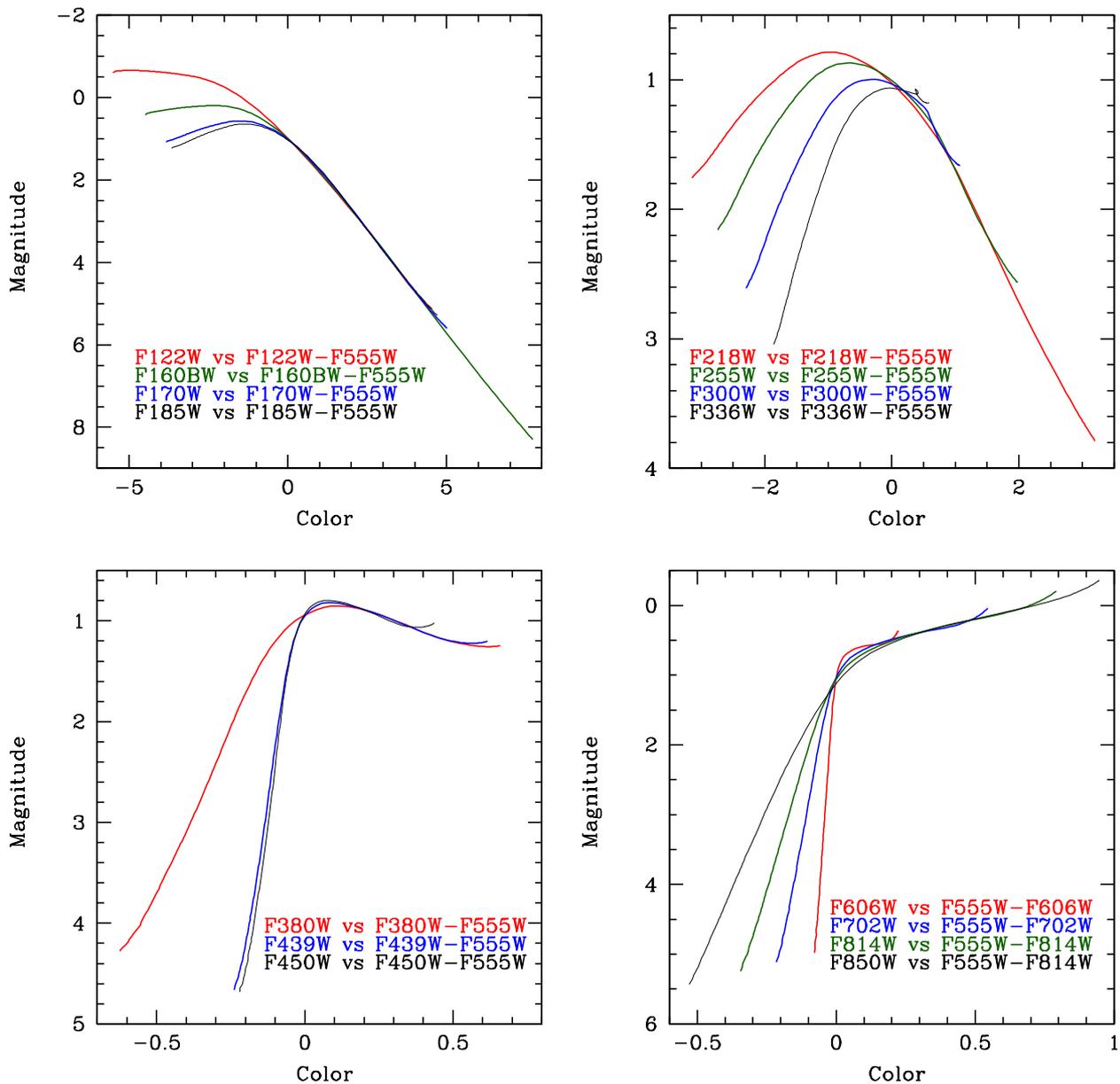}
\caption{ZAHB loci for $Y=0.246$ and $Z=0.001$ for different combinations of HST-WFPC2 filters, as indicated in the insets. F336W is the only filter producing the loop along the ZAHB locus (thin black line in the upper right panel).}
\label{FigZAHBfilters}
\end{figure*}

Is the loop feature only a challenge for F336W, or are other UV filters subject to the same problem?

To answer this question, we applied the same HB models as shown in Fig.~\ref{FigHBTracksUV}--\ref{FigHBTracksUVzoom} to other UV and optical filters in the HST set, using bolometric corrections from \citet{Castelli_Kurucz2004}. The filters included in this analysis are the following: in the far- and near-UV, F122W, F160BW, F170W, F185W, F218W, F255W, F300W, and F380W; in the visible, F439W, F450W, F555W, F606W, F702W, F814W, and F814W. Interestingly, CMDs produced using all of these filters are completely devoid of such a loop feature (Fig.~\ref{FigZAHBfilters}). F300W is noteworthy in this context, as the relevant CMDs appear to attempt to develop a similar loop, but do not quite succeed at the task.

Clearly, then, the loop feature affecting F336W is caused solely by the corresponding bolometric corrections. Over the corresponding temperature region, and as we have mentioned previously, the degeneracy involves a parameter in addition to $Y$ and $t_{\rm HB}$, namely the mass $M_{\rm HB}$, as stars with different $M_{\rm HB}$ values but the same $({\rm F336W}-{\rm F555W})$ can have vastly different absolute magnitudes in F336W (Figs.~\ref{FigHBTracksUV}-\ref{FigHBTracksUVzoom}). Close examination of these figures, along with the synthetic HBs shown in Fig.~\ref{FigSyntheticHBs}, suggests that this triple degeneracy phenomenon can lead to an incorrect reading of the He abundance that may be as high as $\Delta Y=0.02$, in the specific case under study. Note, in particular, that HB stars with $Y=0.246$ starting their evolution close to the loop region will become as bright as the ZAHB locus for $Y=0.266$ by the time they reach the MAHB (see evolutionary tracks for $M_{\rm HB}=0.665$ and $0.652 \, M_\odot$ in Fig.~\ref{FigHBTracksUVzoom}). 

\begin{figure*}
\includegraphics[height=18cm, trim=0.0cm 0cm 0cm 0.0cm]{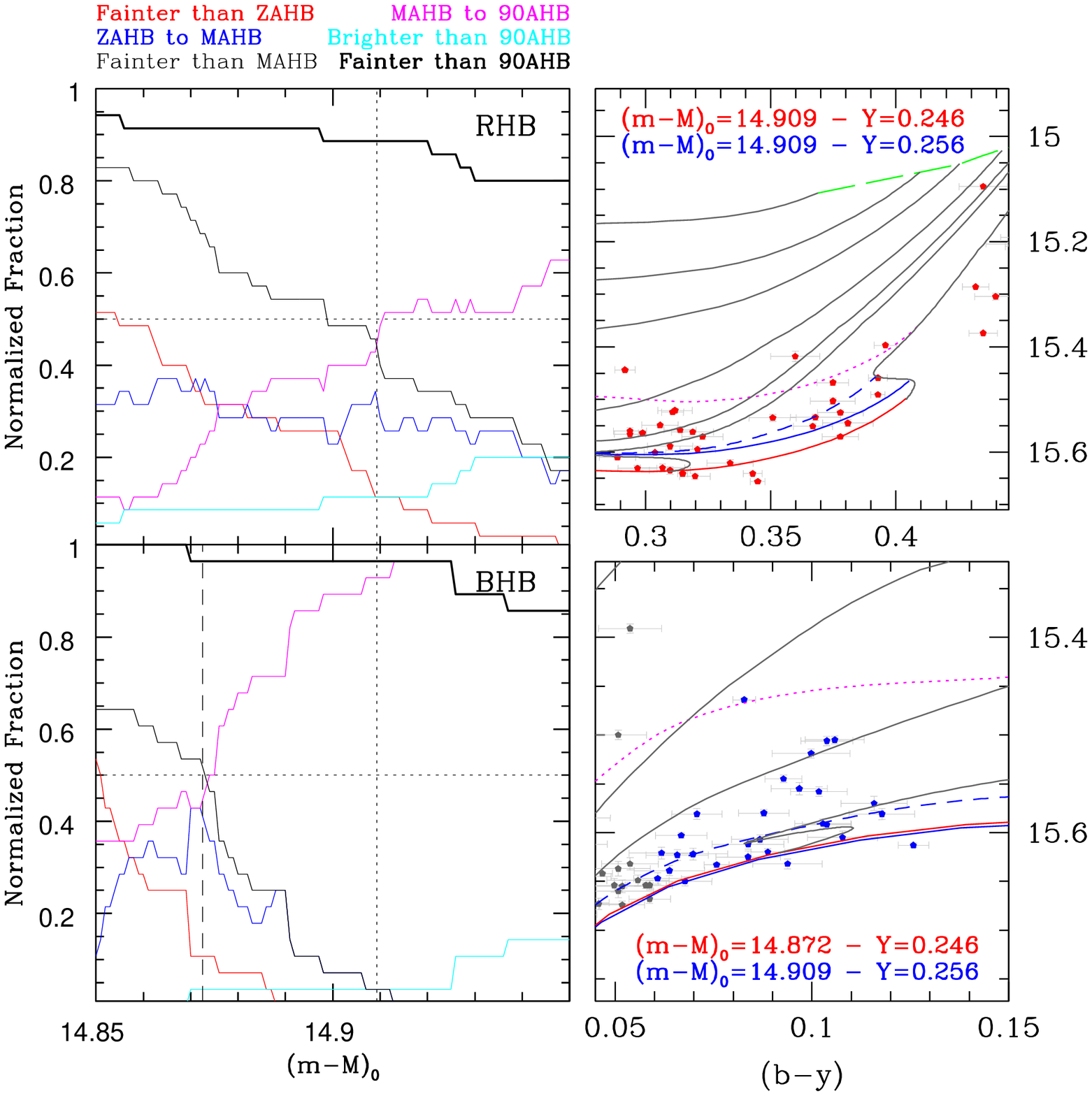}
\caption{Distance modulus determination for the Str\"omgrem filters according to the fraction of stars in each evolutionary stage along the HB, for $Z=0.001$ and $Y=0.246$. The upper and lower panels on the left show the normalized fractions for stars with $0.28<(b-y)<0.44$ (RHB stars) and $0.060<(b-y)<0.135$ (BHB stars), respectively, for filter $y$. Color lines represent the normalized fraction of stars that are (1) fainter than the ZAHB locus (red lines), (2) between the ZAHB and MAHB loci (blue lines), (3) between the MAHB and 90AHB loci (magenta lines), (4) between the 90AHB and TAHB loci (cyan lines), (5) fainter than the MAHB locus (thin black lines), and (6) fainter than the 90AHB locus (thick black lines). Vertical dotted lines represent the distance modulus used in the upper right panel (and Fig.\ref{FigHBTracksStr}), and the vertical dashed line in the lower right panel represents the distance modulus that should be selected if only BHB stars are considered. In the bottom panel the ZAHB locus with $\Delta Y=0.01$ is shifted according to the distance modulus obtained from the RHB population for comparison.}
\label{FigHBDmod}
\end{figure*}

Based on this HB $Y$ test of the presence of He enhancement and internal spreads in GCs, our suggestion would be to avoid relying excessively on the F336W filter over the temperature range $5\,800\lesssim T_{\rm eff}{\rm [K]}\lesssim 8\,300$. Other near- and far-UV filters can be safely used, as can F336W outside this temperature range. Naturally, for temperatures in excess of the Grundahl jump at $T_{\rm eff} \simeq 11\,500$~K, it is necessary to properly take into account the effects of helium sedimentation and radiative levitation of heavier elements \citep{Grundahl_etal1999}, as done (albeit in an approximate way) for NGC~2808 by \citet{Dalessandro_etal2011}.

\subsection{On the $\Delta Y$ level among M3's BHB stars}
\label{SectDistMod}

Following these results, we used the method described in V14 to obtain a new estimate of the difference in (initial) helium content between RHB ($0.28<b-y<0.44$) and BHB ($0.060<b-y<0.135$) stars in M3. While we used the same data in the visual Str\"omgren bands as in C09, here we restrict the bluest color for BHB stars as described in the previous section, and study the variation in the relative proportions of stars distributed in the CMD between the ZAHB, MAHB, 90AHB, and TAHB loci in greater quantitative detail for different adopted distance moduli. We allow for the possibility of a formal solution in which the distance modulus, as inferred on the basis of the BHB stars for an assumed $Y$ value, is different from the one derived using the RHB stars for the same $Y$. Since physically both distance moduli should obviously be the same, any such differences may be indicative of a difference in intrinsic luminosity, hence in the He abundance, between the two HB components. 

The results are shown in Fig.~\ref{FigHBDmod}, where the left panels show the variations in the normalized fraction of stars at different evolutionary stages for RHB stars (upper panel) and BHB stars (bottom panel) as a function of the adopted distance modulus. Since photometric errors can induce variations in these fractions, we considered at the same time the normalized fraction of stars that are (1) fainter than the MAHB (thin black lines), (2) between the MAHB and the 90AHB (magenta lines), and (3) brighter than the 90AHB (cyan lines). The corresponding normalized fractions are expected to be around 50\%, 40\%, and 10\% for each of these three ranges, respectively.

The right panels in the same figure show the CMDs of RHB (red dots) and BHB (blue dots) stars using the distance moduli that best reproduce the expected evolutionary proportions shown in the left panels (dotted and dashed vertical lines, respectively). Theoretical models are shown for $Y=0.246$ (HB tracks in gray lines, ZAHB locus as a red line, MAHB locus as a blue dashed line, 90AHB locus as a dotted magenta line, and TAHB locus as a long green dashed line) and for $Y=0.256$ (ZAHB locus only, shown as a blue line). In the bottom right panel the ZAHB locus for $Y=0.256$ uses the distance modulus obtained for the RHB stars only as a reference locus.

Figure~\ref{FigHBDmod} suggests that the RHB and BHB stellar populations do indeed appear to be best characterized by difference distance moduli. Specifically, the formal distance moduli for the RHB and BHB populations are $14.909^{+0.002}_{-0.006}$ and $14.872\pm0.002$~mag, respectively; more exactly, the difference amounts to 0.037~mag in $y$. Motivated by the HB simulations shown in the top panels of Fig.~\ref{FigSyntheticHBs}, we attribute the few RHB stars that are fainter than the ZAHB locus with $Y=0.246$ as a consequence of photometric errors, of a small population with lower $Y$, and/or of the need of a small increase in the metallicity of PGPUC models. Incidentally, such a formal difference in distance moduli corresponds very closely to what would be expected if the BHB component were more He-rich than the RHB component, by $\Delta Y\approx 0.010\pm0.002$ (where the indicated error bar is the formal value including only the Poissonian component). Assuming this level of He enhancement for the BHB component, the same distance modulus as obtained for the RHB provides an excellent match to the BHB component as well, as shown in the bottom right panel of Fig.~\ref{FigHBDmod}.

\begin{figure}
\includegraphics[width=9.2cm]{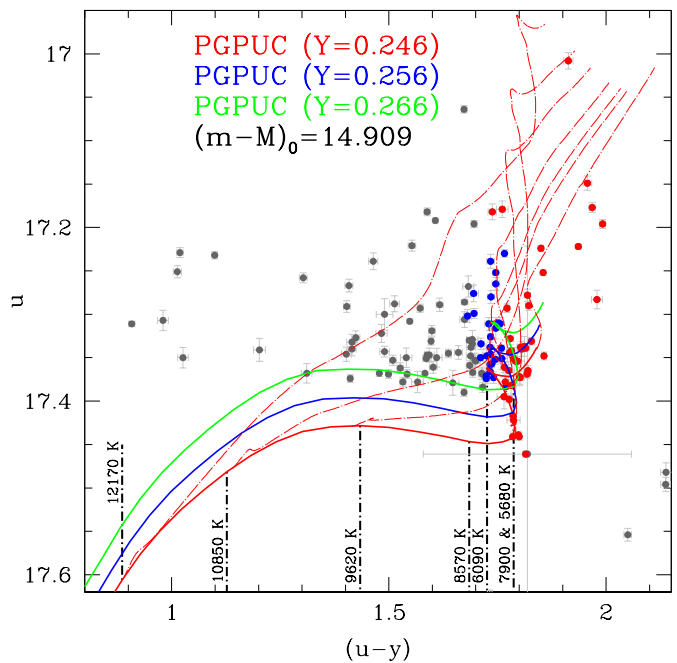}
\caption{M3 HB stars compared to theoretical HB models with $Z=0.001$ in the Str\"omgren $u$ vs. $(u-y)$ plane. Lines have the same meaning as in Fig.~\ref{FigHBTracksStr}. Red and blue points represent the RHB and BHB stars, respectively, that are used to obtain the distance modulus in Fig.~\ref{FigHBDmod}.}
\label{FigM3_uy}
\end{figure}

Unfortunately, this method cannot be as straightforwardly used with the F336W filter because of the difficulties that are brought about by the triple degeneracy effect mentioned previously. This is not only a feature observed in the F336W filter, but also in filters sampling the same wavelength range (see Sect.~\ref{ZAHBloopUband}). In particular, Fig.~\ref{FigM3_uy} shows M3 HB stars in the Str\"omgrem $u$ vs. $(u-y)$ plane with stars color-coded as in Fig.~\ref{FigHBDmod}. Clearly, because RHB and BHB stars overlap at $(u-y)\approx 1.75$, an interpretation of these CMDs is far from being straightforward.

\begin{table*}[]
\centering
\begin{tabular}{|lcrccc|}
\hline
Phase    &$T_{\rm eff}$ range [K]        & $\Delta Y\,\,$  & \% non-variables &  \% total HB & Main diagnostic \\
\hline
RHB+V    &$< 7\,500$                     & $<0.01$  &$32\pm5\%$&$58\pm6\%$ &Str\"omgren (visual) \\
cool BHB &$7\,500\leftrightarrow\,\,\,8\,300$  & $0.01$   &$21\pm4\%$&$13\pm3\%$ &Str\"omgren (visual) \\
BHB      &$8\,300\leftrightarrow\,\,\,9\,300$  & $<0.02$  &$25\pm4\%$&$15\pm3\%$ &WFPC2 (near UV)  \\
hot BHB  &$9\,300\leftrightarrow10\,900$ & $0.02$   &$16\pm3\%$&$10\pm2\%$ &WFPC2 (near UV) \\
eBHB     &$>10\,900$                     & $?\,\,\,\,$&$\,\,\,6\pm2\%$&$\,\,\,4\pm1\%$ & \\
\hline
\end{tabular}
\caption{Amount of He enhancement of M3 BHB stars compared to RHB stars. Fractions are calculated considering only non-variable stars and that the total HB population follows a relation $B:V:R=0.47:0.37:0.16$,  where $B$, $V$, $R$ stand for the number of blue, variable (RR Lyrae), and red HB stars, respectively \citep{Catelan_etal2001}.}
\label{Table1}
\end{table*}

Note, however, that over the color range $0.10 \lesssim({\rm F336W}-{\rm F555W})\lesssim 0.30$, Fig.~\ref{FigHBTracksUV} reveals a lack of HB stars fainter than the ZAHB locus with $Y=0.256$. This can plausibly be associated with an M3 component with a small helium enrichment, that is, with $\Delta Y \approx 0.01$, with $\Delta Y$ possibly increasing by an additional $0.005-0.01$ as the blue limit of this range is approached. Over this color range $17\pm4\%$ of the non-variable HB stars are located, which corresponds to $11\pm2\%$ of the total HB population in M3 \citep[considering RR Lyrae stars;][]{Catelan_etal2001}. Similarly, over the color range $-0.32 \lesssim({\rm F336W}-{\rm F555W})\lesssim 0.10$, the same figure shows a lack of HB stars fainter than the ZAHB locus with $Y=0.266$ that can be associated with an M3 component with a larger helium enrichment ($\Delta Y \approx 0.02$), comprising $16\pm3\%$ of the non-variable HB stars ($10\pm2\%$ of the total HB population). Based on these results, we cannot rule out at present that M3 HB stars that are hotter than the Grundahl jump may have been subject to a higher level of He enhancement than their cooler BHB counterparts~-- but these comprise only about $4\pm1\%$ of the total HB population in M3 (C09).

In Table \ref{Table1} we summarize the level of He enhancement inferred from Figs.~\ref{FigHBTracksStr} and \ref{FigHBTracksUV}, separating the different M3 components according to their temperatures and colors. The number fractions were determined assuming that the non-variable HB stars comprise 63\% of the whole HB populations \citep{Catelan_etal2001}.

Note that, even though multiple populations show different radial distribution in some GCs \citep[e.g.,][]{Lardo_etal2011, Milone_etal2012, Bellini_etal2013} that may be reflected in different BHB-to-RHB number ratios \citep{Catelan_etal2001, Iannicola_etal2009}, we did not observe a large difference in these ratios when the Str\"omgren photometry \citep[with a field of view of $3\farcm75\times3\farcm75$][]{Grundahl_etal1999} was compared with the HST WFPC2 photometry (with a field of view of $2\farcm5\times2\farcm5$), as both sample basically cover the same (central) region of the cluster.

This means that for the range of temperatures covered by the HB $Y$ test, as applied in C09, in V14, and in this paper, the visual bandpasses of the Str\"omgren and Johnson filter systems clearly have a much more straightforward interpretation at temperatures below $\sim8\,300\, {\rm K}$, more directly constraining the level of He enhancement among M3 (and M4; see V14) stars alike at $\Delta Y \lesssim 0.01$, which confirms the C09 results for the bulk of the HB stars in the cluster. At higher temperatures, however, these same bandpasses lose sensitivity, and it becomes necessary to resort to filters covering the UV regime, some of which (F336W in particular) lead to CMDs that are difficult to interpret at lower temperatures. Using data obtained with the WFPC F336W, we were able to successfully identify small populations of hotter HB stars in M3 that very likely have slightly enhanced helium, reaching up to 0.02 in $\Delta Y$. This supports similar conclusions reached by D13 for this hot HB component.

\section{On the ZAHB loop in U bands of additional filter sets}
\label{ZAHBloopUband}

Since this {\em loop} feature is observed in the HST/WFPC2 F336W vs. $({\rm F336W}-{\rm F555W})$ plane (Fig.~\ref{FigHBTracksUV}) and in the Str\"omgrem $u$ vs. $(u-y)$ plane (Fig.~\ref{FigM3_uy}), we now address its possible presence in two other filter systems equipped with bandpasses that cover a similar wavelength regime as F336W or Str\"omgren $u$, namely Johnson-Cousins \citep{hj55,ac76} and  Sloan \citep{mf96}. However, we emphasize that these are not the only filter combinations for which loop-like features are predicted by our models. Loops along the ZAHB are found in most of the cases where filters covering the same wavelength regime are used to define color indices.

\begin{figure*}
\includegraphics[height=18.5cm]{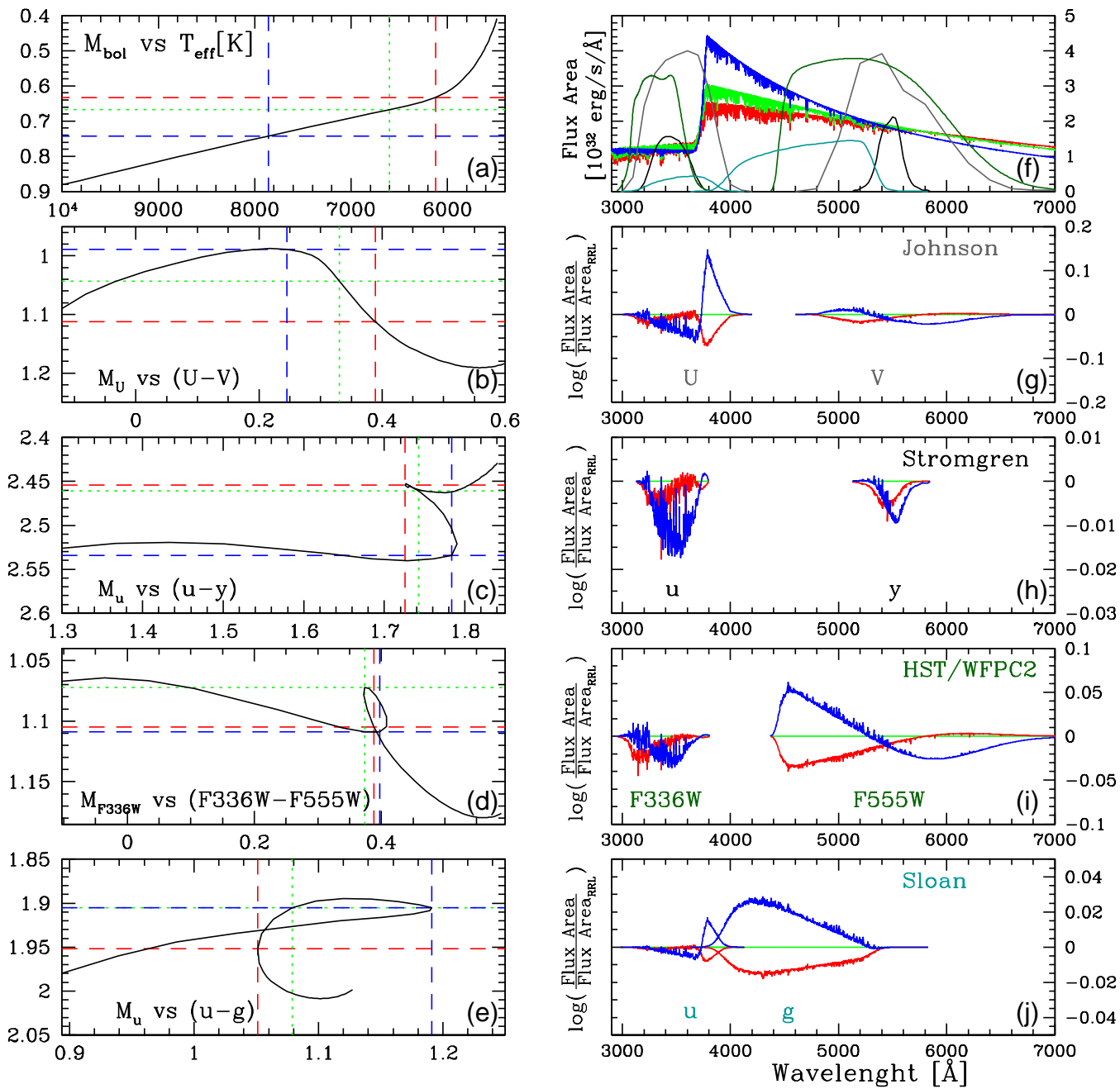}
\caption{{\bf Panels a to e} show a ZAHB locus for $Y=0.246$ and $Z=0.001$ (black lines) in the theoretical plane ($M_{bol}$ vs $T_{\rm eff}$) and in four UV-bands (Johnson-Cousin $U$ vs. $U-V$; Str\"omgrem $u$ vs. $(u-y)$; HST/WFPC2 $F336W$ vs. $({\rm F336W}-{\rm F555W})$; Sloan $u$ vs. $u-g$). Dashed and dotted lines indicate the location at the ZAHB locus of the following temperatures: $6\,125\,{\rm K}$ (red line), $6\,600\,{\rm K}$ (green line), and $7\,800\,{\rm K}$ (blue line). {\bf Panel f} shows the spectra of stars at the ZAHB level with the properties ($T_{\rm eff}$, $\log g$) of dashed/dotted lines computed with ATLAS9 and SYNTHE packages. Transmission curves of the filters used in left panels  are also shown: Johnson $U-V$ (gray lines), Str\"omgren $u-y$ (black lines), HST/WFPC2 F336W$-$F555W (dark green lines), and Sloan $u-g$ (cyan lines). Transmission curves are multiplied arbitrarily by 4. To compare these spectra, fluxes are multiplied by the surface area of the respective star. {\bf Panels g to j} show the differential convolved spectra depending on the transmission curves compared to the $6600\, {\rm K}$ spectra.}
\label{FigTeo_Ubands}
\end{figure*}

In Fig. \ref{FigTeo_Ubands} we show the ZAHB loci for $Y=0.246$, $Z=0.001$ as black lines in the theoretical plane (panel a) and in empirical planes associated with these four different filter systems, namely: Johnson-Cousins $U$ vs. $U-V$ (panel b); Str\"omgren $u$ vs. $u-y$ (panel c); HST/WFPC2 F336W vs. ${\rm F336W}-{\rm F555W}$ (panel d); Sloan $u$ vs. $u-g$ (panel e). To compare the range of effective temperatures restricting the {\em loop} in the F336W vs. ${\rm F336W}-{\rm F555W}$ plane with other sets of filters, we show lines representing $T_{\rm eff}=6\,125\,{\rm K}$ (red dashed lines), $6\,600\,{\rm K}$ (green dotted lines) and $7\,800\,{\rm K}$ (blue dashed line) in these panels. As can be seen, the Sloan $u$ vs. $u-g$ plane also shows a {\em loop} feature along the ZAHB locus, but in this case it is even more impressive, extending as it does up to $\sim9\,500\,{\rm K}$. For the Johnson-Cousins system, the $U-V$ color index monotonically decreases along the ZAHB as $T_{\rm eff}$ increases, which makes these filters immune to the aforementioned triple degeneracy effect, and thus affords a cleaner separation between BHB and RHB  stars than is the case with the other three filter systems that we studied.

What is the origin of this {\em loop} phenomenon, and why is it seen more prominently in some filter systems than in others? To answer these questions, we computed stellar atmosphere models and synthetic spectra for three stars along the {\em loop}, using the ATLAS9 and SYNTHE packages \citep{Castelli_Kurucz2004,Sbordone_etal2004}. We assumed a chemical composition given by $Z=0.001$ and $[\alpha/{\rm Fe}]=+0.4$. These three stars have physical parameters corresponding to the intersection between the different lines and the ZAHB sequence as shown in Fig.~\ref{FigTeo_Ubands}a. Thus, they correspond to an RHB star with $T_{\rm eff}=6\,125\,{\rm K}$, $\log g=2.71$, $M_{HB}=0.663 \, M_\odot$ (red lines, corresponding to the start of the loop in panel d); an RR Lyrae with $T_{\rm eff}=6\,600\,{\rm K}$, $\log g=2.85$, $M_{HB}=0.652 \, M_\odot$ (green lines, middle of the loop in panel d), and a BHB star with $T_{\rm eff}=7\,800\,{\rm K}$, $\log g=3.15$, $M_{HB}=0.635 \, M_\odot$ (blue line, end of the loop in the same panel). The resulting synthetic spectra are shown in panel f of the same figure. 

To make the comparison between these spectra more straighforward, we then multiplied the derived fluxes by the (different) surface areas of these stars, thus placing the results in luminosity units. The monochromatic luminosities thus derived were then multiplied by the transmission curves of the different filters used in our analysis, which are also displayed in Fig.~\ref{FigTeo_Ubands}f (Johnson-Cousins system: gray lines; Str\"omgrem system: black lines; HST/WFPC2 system: dark green lines; Sloan system: cyan lines)\footnote{Transmission curves were downloaded from the Spanish Virtual Observatory webpage at \url{http://svo2.cab.inta-csic.es/svo/theory/fps3/index.php} .}. The results for the early- and late-loop stars were then analyzed by their ratios with respect to the middle-loop star. The results for the different filter systems are shown in panels g through j in the same figure, each corresponding to a different filter system. 

Before proceeding with the analysis, we recall that the area below the spectrum in the wavelength range covered by a given filter $X$ is related to the bolometric correction ($BC_X$) in that same filter. The absolute magnitude in the same filter is then computed by means of the relation $M_X=M_{\rm bol}-BC_X$. Therefore, higher $BC_X$ values will imply brighter stars in the $X$ filter, and vice versa. Similarly, the difference in areas covered by two different filters $X$ and $Y$ will be related to the color, in such a way that $X-Y=BC_Y-BC_X$. A star may thus appear bright in a filter $X$ either because it is intrinsically very bright (small $M_{\rm bol}$) or/and because it has a large bolometric correction $BC_X$. Keeping these concepts in mind, we can now examine each filter system in turn.

In the Johnson-Cousins CMD (panel b), the end of the loop is brighter in $U$ than the beginning of the loop, even though the opposite occurs for $M_{\rm bol}$ (panel a). Panel g in the same figure shows that this is due to the significant contribution of the Balmer jump, which leads to a high $BC_U$ value for this star, but not for its cooler counterparts along the loop.

The behavior for the Str\"omgren CMD (panel c) is different: in this case, similarly to what occurs with $M_{\rm bol}$, the star is brighter in $u$ at earlier stages of the loop (i.e., lower temperatures) than it is towards its end because the Str\"omgren $u$ filter is practically unaffected by the Balmer jump. The behavior of the $(u-y)$ color, however, is non-monotonic along the loop, which can be understood from panel h by noting that the difference in areas (and thus in the $BC$ values) decreases slightly at higher temperatures, implying a relative brightening of the star in $y$ compared to $u$, which causes the star to appear redder toward the end of the loop than at earlier stages.

For HST/WFPC2 (panel d), the magnitude $M_{\rm F336W}$ is dominated by $BC_{\rm F336W}$ as in the Johnson-Cousins filters, but since the F336W does not reach the Balmer jump, $BC_{\rm F336W}$ increases from lower temperatures to $\sim6\,600\,{\rm K}$ and then decreases for higher temperatures, producing as a result the characteristic up-and-down morphology of the loop. In turn, the non-monotonic behavior of the ${\rm F336W}-{\rm F555W}$ color arises because the differences between the areas corresponding to the blue and red convolved spectra are similar at the beginning and end of the loop, thus giving rise to similar ${\rm F336W}-{\rm F555W}$ colors at these two points~-- meaning that they close the loop.

Finally, the behavior that is seen in the Sloan CMD (panel e) can be understood from the fact that the Sloan $u$ filter is sensitive to the Balmer jump, similar to what is seen in the Johnson-Cousins $U$ filter (but not in Str\"omgren $u$ or HST/WFPC2 F336W). As a consequence, and as was also seen for $U$, the Sloan $u$ magnitude variation across the same loop region as mapped in panel d is also monotonic. For the $u-g$ colors, it is clear that there is a larger difference in area between the $u$ and $g$ filter regimes in the case of the blue convolved spectrum than in the case of the red convolved spectrum. For this reason, the color is redder at the high-temperature end of the loop than it is at its cool and intermediate parts. Evidence for this effect is indeed present in the M3 SDSS photometry obtained by \citet[][see their Fig. 10a]{An_etal2008}.

In summary, the {\em loop} feature along the ZAHB locus that is seen in some but not all near-UV filters is not caused by any specific set of absorption lines, but is largely controlled by the extent to which any given filter may cover the Balmer jump, as well as to the way in which the different filters sample the stellar spectra over the relevant temperature region. 

To close, we provide for convenience the temperature range over which the mixing of ZAHB populations can be avoided in CMDs obtained on the basis of different filter systems for a chemical composition similar to that of M3. a)~In the Johnson-Cousins $M_U$ vs. $U-V$ CMD, there are no restrictions. b)~In the Str\"omgren $u$ vs. $(u-y)$ CMD, the mix can be avoided for stars with $T_{\rm eff}\gtrsim8\,700$ K ($u-y\lesssim1.65$). c)~In the HST/WFPC2 F336W vs. (F336W$-$F555W) CMD, no mixing is seen for ZAHB stars with $T_{\rm eff}\gtrsim8\,300$ K $({\rm F336W-F555W}\lesssim 0.34$). d)~In the Sloan $u$ vs. $u-g$ diagram, the region without population mixing corresponds to $T_{\rm eff}\gtrsim9\,500$ K ($u-g\lesssim1.02$). Proper analysis of the $T_{\rm eff}\gtrsim11\,500$~K regime, however, requires computing bolometric corrections that properly take the effect of metal levitation and He sedimentation into account \citep{Grundahl_etal1999,Behr2003}.

\section{Conclusions}

In this paper, we have revisited the problem of the level in He enhancement of M3 HB stars using both HST data extending to the near UV (F336W filter) and ground-based Str\"omgren data in the visible ($b,y$ filters). To interpret the data, we computed a large set of HB models, including different levels of He enhancement, for a chemical composition appropriate to M3.

Our results reveal that CMDs based on the F336W filter develop a unique {\em loop} feature (observed also in several similar filters), at temperatures ranging from the RR Lyrae instability strip to the cool component of the BHB, which leads to a triple degeneracy in CMD position, in the sense that multiple combinations of $Y$, stellar mass, and evolutionary age since the ZAHB can lead to stars occupying very closely the same spot on the CMD. We argue that this phenomenon can affect the interpretation of CMDs that use F336W, potentially leading, for M3, to an incorrect reading of the level of He enhancement among the cool BHB stars by as much as $\Delta Y \approx 0.02$, over the temperature range $8\,300\gtrsim T_{\rm eff}[{\rm K}]\gtrsim5800$ ($0.34 \lesssim {\rm F336W}-{\rm F555W} \lesssim 0.45$). Other CMDs based on both bluer and redder HST filters, as well as those based on the Str\"omgren visual filters, are immune to this effect and can thus be more straightforwardly interpreted, particularly for $T_{\rm eff}\lesssim8\,300$~K $(b-y \gtrsim0.06)$. We suspect that this is the main reason why D13 (who used F336W in their analysis) invoked a higher level of He enhancement among the cooler BHB stars in M3 than did C09 (who used the Str\"omgren visual filters). 

Thus, we estimate a $\Delta Y\approx 0.01$ for BHB stars in M3 with $T_{\rm eff}\lesssim8\,300$~K, which comprise $13\pm3\%$ of the total population of HB stars in M3. For temperatures ranging from $\approx8\,300$~K to $\approx9\,300$~K, a $\Delta Y\approx 0.01-0.02$ is estimated possibly increasing with increasing $T_{\rm eff}$, corresponding to $15\pm3\%$ of the total population of HB stars in M3. A low level of He enhancement among such stars qualitatively agrees with the conclusion of C09, although quantitatively, the level of He enhancement found here, particularly for the hotter stars, is higher than favored in C09. The latter study, based as it was solely on visual bandpasses, lacked sensitivity to the hotter components of the blue HB, which mostly likely present a higher level of He enhancement, as revealed in the near UV. Conversely, for hotter HB stars, our results agree better with those of D13. In particular, our analysis suggests a higher level of He enhancement, $\Delta Y\approx 0.02$, over the temperature range $9\,300\lesssim T_{\rm eff}[{\rm K}]\lesssim10\,900$, corresponding to $10\pm2\%$ of M3 HB stars. Last but not least, a minority population ($4\pm1\%$ of the total HB population) of hot HB stars ($T_{\rm eff}\gtrsim10\,900$~K, that is, hotter than the Grundahl jump) with an even higher level of He enhancement cannot be excluded at present, but an analysis of these stars is severely hampered by the need to take gravitational diffusion and radiative levitation properly into account.
  
To gain insight into the {\em loop} phenomenon, we finally computed stellar spectra along the ZAHB locus over the relevant temperature range and convolved them with near-UV filters around the $U$ region provided in different filter systems (Johnson-Cousins, Str\"omgren, HST/WFPC2, Sloan). We showed that the loop behavior can be understood by the way in which different filters sample the stellar spectra, particularly around the Balmer jump. We also found that CMDs based on the Johnson $U$ filter do not develop the {\em loop}, and thus afford the cleanest interpretation of the empirical data. Conversely, the {\em loop} becomes most pronounced when using the Sloan $u$ filter.

\begin{acknowledgements}
We thank the anonymous referee for the comments that have greatly helped improve the presentation of our results. We also thank Emanuele Dalessandro for sending us his HST photometry for M3, and Luca Sbordone and Fiorella Castelli for their expert advice in the use of the ATLAS9 and SYNTHE packages. Support for A.A.R.V., M.C., J.A.-G., R.C.R., and S.A. is provided by the Ministry for the Economy, Development, and Tourism's Programa Iniciativa Cient\'{i}fica Milenio through grant IC\,120009, awarded to the Millennium Institute of Astrophysics (MAS); by Proyecto Basal PFB-06/2007; and by Proyecto FONDECYT Regular \#1141141. A.A.R.V., J.A.-G., and  R.C.R. acknowledge support through FONDECYT postdoctoral grants \#3140575, 3130552, and 3130320, respectively. J.A.-G. acknowledge support by the FIC-R Fund, allocated to the project 30321072. This work was done (in part) using the Geryon computer at the Center for Astro-Engineering UC, part of the Basal PFB-06 project, which received additional funding from QUIMAL 130008 and Fondequip AIC-57 grants for upgrades. Additional support for this project is provided by CONICYT's PCI program through grant DPI20140066.
\end{acknowledgements}

\bibliographystyle{aa}
\bibliography{avalcarceM3_2016}

\end{document}